\documentclass[aps,pre,a4paper,twocolumn,amssymb,amsfonts]{revtex4-1} 
\pdfoutput=1
\usepackage{mathptmx} 
\usepackage{bm}
\usepackage{graphicx}
\usepackage{color}
\usepackage{hyperref}
\hypersetup{
pdfborder={0 0 0}, 
colorlinks=true, 
linkcolor=blue,
citecolor=blue,
urlcolor=blue
}

\begin{document}

\title{Comment on ``Reconnection of quantized vortex filaments and the Kolmogorov spectrum''}

\author{R. H\"anninen}
\affiliation{Low Temperature Laboratory, Department of Applied Physics, Aalto University, P.O. Box 15100, FI-00076 AALTO, Finland}

\date{\today}

%
\begin{abstract}
In this Comment we would like to emphasize that in \href{http://dx.doi.org/10.1103/PhysRevB.90.104506}{Phys. Rev. B {\bf 90}, 104506 (2014)} the calculated energy spectrum takes into account only the small interaction (cross) term and, additionally, this term is only calculated at the instant when the two vortices reconnect. The majority of the kinetic energy is contained in the self-energy term which has a characteristic spectrum of $1/k$. If this, and the additional average over time, is taken into account the suggested Kolmogorov type $k^{-5/3}$ spectrum is likely not visible in the kinetic energy spectrum which contains both terms. Therefore, we find the suggestion misleading that the Kolmogorov spectrum in superfluids arises from the reconnection of vortices.    
\end{abstract}
%
\maketitle

\section{Introduction}\label{s.intro}

In classical turbulent fluids, the famous Kolmogorov energy spectrum $E(k)\propto k^{-5/3}$ is one of the most well-known indications of the energy flux through different length scales, until dissipation becomes important at the smallest scales. In superfluids the viscosity is zero and due to quantization of the vorticity a similar Kolmogorov-type cascade is only possible at scales larger than the intervortex distance. A classical-like spectrum is typically, but not always, explained by the formation of vortex bundles of different size that can mimic classical turbulence at scales larger than the intervortex spacing [\onlinecite{NemirovskiiPR2013}]. In Ref.~[\onlinecite{SKN}] a $k^{-5/3}$ spectrum is suggested to result from the reconnection of two vortices. We argue that this conclusion is rather misleading due to the omitted terms, and also because a proper time average is omitted.   

\section{Energy spectrum from line vortices}\label{s.spectrum}

At scales larger than the vortex core, quantized vortices can be approximated using the vortex filament model where the superfluid velocity is given by the Biot-Savart law. The (incompressible) kinetic energy spectrum (averaged over the solid angle for the ${\bf k}$ vector) can be written in the form given in Ref.~[\onlinecite{SKN}]:
\begin{equation}
E(k)\! = \!\frac{\rho_s\kappa^2}{(2\pi)^2}\!\!\oint\!\!\oint \hat{\bf s}'(\xi_1)\cdot \hat{\bf s}'(\xi_2) \frac{\sin(k|{\bf s}(\xi_1)\!-\!{\bf s}(\xi_2)|)}{k|{\bf s}(\xi_1)\!-\!{\bf s}(\xi_2)|}d\xi_1d\xi_2, \label{e.Ek}
\end{equation}
where ${\bf s}(\xi)$ describes the vortex location with the unit tangent $\hat{\bf s}'(\xi)$. In the equation both line integrals sweep the whole vortex tangle [in Ref.~[\onlinecite{SKN}] the two vortices (1) and (2)], described by the arc length $\xi$. Here one needs to assume that the vortices form closed loops. In Ref.~[\onlinecite{SKN}] the author states that ``I calculated only the interaction energy between the approaching parts of different lines; the self-energy in the vicinity of the point of contact vanishes, since the lines are antiparallel.'' This means that the calculated energy only contains the integrand where $\xi_1$ and $\xi_2$ belong to the different vortices. However, the self-energy term where $\xi_1$ and $\xi_2$ belong to the same vortex is nonzero everywhere, the dominant part coming when $\xi_1 \approx \xi_2$, also near the contact point [\onlinecite{note1}]. This contribution to the energy is generally the most important one but is left totally unnoticed in Ref.~[\onlinecite{SKN}]. 
The self-energy term is present in the experiments which trace the total energy distribution and it has been included in the spectrum when analyzing the previous vortex filament simulations.
It produces the known result for the energy spectrum at length scales smaller than the average intervortex distance, giving $E(k)\propto 1/k$, i.e., loosely speaking it results from the spectrum of a straight vortex which generates a velocity field ${\bf v}_{\rm s}=\kappa/(2\pi{r})\hat{\bm \phi}$ around its core. The amplitude of the energy spectrum at large $k$ is simply given by the vortex length. This is also true even if the vortices support Kelvin waves (helical distortions) [\onlinecite{NiklasJLTP2013,NemirovskiiJLTP2013,note2}]. We have additionally analyzed the total energy spectrum for reconnecting vortex rings, the situation described and otherwise analyzed in Ref.~[\onlinecite{HanninenPRB2013}], and only found the $1/k$ spectrum at length scales smaller than the initial ring radius. In many early numerical simulations with the vortex filament model this asymptotic limit is poorly derived [\onlinecite{KivotidesPRL2001,ArakiPRL2002,KondaurovaJLTP2005}].

In the vortex filament simulations the $k^{-5/3}$ spectrum is realized on scales larger than the intervortex distance, not around it as stated in Ref.~[\onlinecite{SKN}]. The only simulations where the Kolmogorov-like spectrum extends to scales smaller than the intervortex distance are Gross-Pitaevskii calculations [\onlinecite{JepezPRL2009}]. There the big question is the effect of the compressibility and the existence of sound waves, which complicate the analysis. Even though this spectrum might be an indication of the Kelvin-wave cascade, a more plausible explanation is given in Ref.~[\onlinecite{LvovPRLcomment}], and references therein. 

Since the integrand of the self-energy term brings a similar contribution to the energy spectrum as the integrand of the interaction term at the contact point, one may easily estimate that the interaction term, after integration, is at least $a/L$ times smaller than the self-energy term, where $a$ is the curvature radius at the tip and $L$ is the total length of the vortices which here can be chosen to be of the same order as the intervortex distance. One should expect that $a/L \ll 1$.  Additionally, since the steady state energy spectrum is an average over time, one should average over a time window where the minimum distance between the two tips varies between 0 and intervortex distance. Since this minimum distance behaves like $d_{\rm min} = A\sqrt{|t-t_{\rm rec}|}$, where $t_{\rm rec}$ is the instant of reconnection, the time window where the interaction term is dominant is only a small fraction around $t_{\rm rec}$ of the total time window. This is a further reason that the $k^{-5/3}$ spectrum from the interaction term is likely to be nonresolvable and masked by the dominant $1/k$ spectrum originating from the self-energy term.

Finally we note that one should be careful when using Eq.~(\ref{e.Ek}) to calculate the energy spectrum for vortex configurations that are not closed loops, as in the example of Ref.~[\onlinecite{SKN}]. This gives results which are erroneous at scales around the system size and larger. Here the system size can be taken as the intervortex distance $\delta \approx 1$ (the author is using dimensionless units). 

\section{Conclusions}\label{s.conc}

The search for the $k^{-5/3}$ spectrum in superfluids originates mainly from the many similarities found between classical fluids and superfluids. Currently the Kolmogorov-type spectrum is only supported by a few numerical simulations illustrating that at length scales larger than the intervortex distance the superfluid mimics classical fluids. A proper explanation is still missing. At length scales smaller than the intervortex distance the Kelvin-wave cascade has been suggested to produce the $k^{-5/3}$ spectrum for the energy [\onlinecite{LvovKWC}]. However, also in this case the spectrum is not for the total kinetic energy, but only for the energy related to the Kelvin waves. If one takes into account the spectrum from the straight vortex one realizes that the total spectrum will be much closer to $1/k$, as emphasized e.g., in Refs.~[\onlinecite{NiklasJLTP2013,NemirovskiiJLTP2013}]. Here, in Ref.~[\onlinecite{SKN}], the suggested $k^{-5/3}$ spectrum resulting from the shape of the reconnection kinks at the instant when two antiparallel vortices reconnect at one point, is also likely to be dominated by the self-energy term that was omitted in the analysis.   

\begin{acknowledgments}
This work is supported by the Academy of Finland (Grant No. 218211). I would like to acknowledge M. Krusius and N. Hie\-ta\-la for 
their comments and suggestions for this comment.  
\end{acknowledgments}


\begin{thebibliography}{9}
\bibitem{note1}
The integrand of the self-energy term at the contact (reconnection) point does not vanish, rather it is the integrand of the interaction energy which exactly cancels the self-terms, but this occurs only at the contact point and does not change the general conclusions presented here or in the original article.
\bibitem{note2}
The amplitude of the energy spectrum at large $k$ is given by the vortex length, provided that the amplitude of the Kelvin waves becomes smaller than their wavelength at high enough $k$-values. This is true for any realistic configuration where the vortex line length is finite. \newline
\end{thebibliography}
\end{document}